\documentclass[twocolumn,tighten]{aastex63}
\usepackage{amsmath}
\usepackage{verbatim}
\usepackage{booktabs}
\usepackage{xcolor}

\shortauthors{Gupta et al.}

\graphicspath{{./}{figures/}}
\def\chandra {{\it Chandra}~}

\def\xmm {{\it XMM-Newton}~}

\def\suzaku {{\it Suzaku}~}

\def\offii {{\it Off-field~2}~}
\def\offiin {{\it Off-field~2}}
\def\offiii {{\it Off-field~3}~}
\def\offiiin {{\it Off-field~3}}
\def\offiv {{\it Off-field~4}~}
\def\offivn {{\it Off-field~4}}
\def\offv {{\it Off-field~5}~}
\def\offvn {{\it Off-field~5}}

\def\oi{{{\rm O}\,{\sc i}~}}

\def\ovii{{{\rm O}\,{\sc vii}~}}
\def\oviin{{{\rm O}\,{\sc vii}}}
\def\oviii{{{\rm O}\,{\sc viii}~}}
\def\oviiin{{{\rm O}\,{\sc viii}}}
\def\neix{{{\rm Ne}\,{\sc ix}~}}

\def\mgxi{{{\rm Mg}\,{\sc xi}~}}
\def\mgxin{{{\rm Mg}\,{\sc xi}}}

\def\mrk {{\it Mrk~509}~}
\def\mrkn {{\it Mrk~509}}

\def\arkn {{\it Ark~564}}

\def\ngcn {{\it NGC~3783}}
\def\h {{\it H2106-099}~}

\begin{document}

\title{Super-virial temperature or Neon overabundance?: \suzaku observations of the Milky Way circumgalactic Medium}

\correspondingauthor{Anjali Gupta}
\email{agupta1@cscc.edu}

\author{Anjali Gupta}
\affiliation{Columbus State Community College, 550 E Spring St., Columbus, OH 43215, USA} 
\affiliation{Department of Astronomy, The Ohio State University, 140 West 18th Avenue, Columbus, OH 43210, USA} 

\author{Joshua Kingsbury}
\affiliation{Columbus State Community College, 550 E Spring St., Columbus, OH 43215, USA} 

\author{Smita Mathur}
\affiliation{Department of Astronomy, The Ohio State University, 140 West 18th Avenue, Columbus, OH 43210, USA} 
\affiliation{Center for Cosmology and Astroparticle Physics, 191 West Woodruff Avenue, Columbus, OH 43210, USA}

\author{Sanskriti Das}
\affiliation{Department of Astronomy, The Ohio State University, 140 West 18th Avenue, Columbus, OH 43210, USA} 

\author{Massimiliano Galeazzi}
\affiliation{Physics Department, University of Miami, Coral Gables, FL 33155, USA} 

\author{Yair Krongold}
\affiliation{Instituto de Astronomia, Universidad Nacional Autonoma de Mexico, 04510 Mexico City, Mexico} 

\author{Fabrizio Nicastro}
\affiliation{Observatorio Astronomico di Roma-INAF, Via di Frascati 33, 1-00040 Monte Porzio Catone, RM, Italy}

\begin{abstract}
We analyzed \suzaku and \chandra observations of the soft diffuse X-ray background toward four sightlines with the goal of characterizing the X-ray emission from the Milky Way circumgalactic medium (CGM). We identified two thermal components of the CGM, one at a uniform temperature of $\rm kT = 0.176\pm0.008~keV$ and the other at temperatures ranging between $\rm kT = 0.65-0.90~keV$. The uniform lower temperature component is consistent with the Galaxy's virial temperature ($\rm \sim 10^{6}~K$). The temperatures of the hotter components are similar to that recently discovered ($\rm \sim 10^7~$K; Das et al.) in the sightline to blazar $1ES~1553+113$, passing close to the Fermi bubble. Alternatively, the spectra can be described by just one lower-temperature component with super-solar Neon abundance, once again similar to that found in the $1ES~1553+113$ sightline. The additional hot component or the overabundance of Ne is required at a significance of $>4\sigma$, but we cannot distinguish between the two possibilities. These results show that the super-virial temperature gas or an enhanced Ne abundance in the warm-hot gas in the CGM is widespread, and  these are not necessarily related to the Fermi bubble.
\end{abstract}


\section{Introduction}

%%%The circum-galactic medium (CGM) is believed to be multi-phase at temperatures $\rm 10^{5}-10^{7}~K$ \citep{Cen1999}. The gas gets to these temperatures by shock heating while accreting from the intergalactic medium (IGM) and/or receiving the feedback from outflows (star-formation and/or AGN induced). The CGM may help sustain and regulate star-formation and may harbor the largest baryon and metal reservoir \citep{White1978, Stinson2012, Oppenheimer2016, Nelson2018}

The circum-galactic medium (CGM) is an important component of a
spiral galaxy. The CGM is defined as the gaseous medium surrounding the stellar disk of a galaxy, extended out to its virial radius. It serves as a gas reservoir with accretion from
the intergalactic medium (IGM) and outflows from the stellar disk (star-formation and/or AGN
induced). Some of this material may recycle back into the disk of the galaxy, while some may
stay in the diffuse CGM. This helps to regulate the formation and evolution of a galaxy. The CGM is also believed to contain more baryonic mass than the entire stellar disk, and also most of the metals produced by stars \citep{White1978, Oppenheimer2016, Nelson2018}. Thus the CGM may harbor the largest galactic baryon and metal reservoir. The CGM is predicted to be predominantly warm-hot: $\rm T \approx 10^{5}-10^{7} ~K$, with most of the baryons in the hotter $\rm T \approx 10^{6}-10^{7} ~K$ phase \citep{Stinson2012}.

The higher temperature ($\rm \ge 10^{6}~K$) phases of the CGM can only be probed by soft X-ray observations, particularly the oxygen transitions of \ovii and \oviiin, both in absorption and emission. Because of our special vantage point, our own Milky Way provides a unique opportunity to probe the CGM of a spiral galaxy. Absorption lines due to \ovii and \oviii at redshift zero, from the Milky Way CGM, have been detected toward extragalactic sight-lines by \chandra and \xmm \citep{Nicastro2002, Wang2005, Williams2005, Williams2006a, Williams2006b, Williams2007, Gupta2012, Gupta2014, Fang2015}. The line ratios of \oviin/\oviii constrain the absorbing gas temperature in the range of $\rm 0.1-0.2~keV~(1.2-2.5 \times 10^{6}~K)$, assuming both absorption lines arise in the same plasma \citep{Gupta2012}.

Various broad band X-ray observations have revealed an extensive soft ($\rm \sim 0.1-1.0~ keV$) diffuse X-ray background (SDXB; \cite{Snowden1998, Snowden2000}). The ``shadow observations," in which the X-ray emission towards a molecular cloud at a known distance is compared with a nearby line of sight with low absorption, show that there is a significant contribution from the Milky Way CGM to the SDXB \citep{Smith2007, Galeazzi2007, Gupta2009, Henley2015}. Several studies have attempted to measure the Galactic CGM contribution to the SDXB using \xmm and \suzaku observations of empty fields (with no bright source in the field-of-view)  \citep{Yoshino2009, Henley2010, Henley2013, Nakashima2018}. All these X-ray emission studies of the warm-hot gas in the Galactic halo\footnote{Different fields have traditionally used different nomenclature to describe the gas filling the Galactic halo. We will therefore use the terms "CGM" and "halo" interchangeably.} have shown that the gas temperature is fairly constant across the sky $\rm 0.15-0.21~keV~(1.8-2.4 \times 10^{6}~K)$ and is comparable to the absorbing gas temperature. 

Although X-ray emission and absorption observations established the presence of the warm-hot gas in the Galactic halo, the extent, density and mass of this warm-hot gas has been a matter of debate. Determining the density ($\rm n_{e}$) and the path-length (R) of the warm-hot gas is critical for estimating its total baryonic mass. However, this is difficult in part due to the insufficiency of current X-ray gratings to resolve the absorption lines, and also because of the degeneracy involved in using absorption or emission studies alone. The strength of an absorption line depends on the ionic column density, which in turn depends on the equivalent hydrogen density  ($\rm N_{\rm H}=\int{n_{H} dr}$) of the intervening gas, while the emission measure (EM) is sensitive to the square of the electron number density ($\rm EM = \int{n_{\rm e}^2dr}$). Combining absorption and emission measurements breaks the degeneracy and provides constraints on the path-length and density of the absorbing/emitting plasma.

In our previous work \citep{Gupta2012}, comparing the absorption and emission values averaged over the whole sky, we found that there is a huge reservoir of ionized gas in the Galactic halo, with a mass of over $\rm \sim 6 \times 10^{10}~M_{\odot}$ and a radius of over $\rm 100~ kpc$. However, shadow observations and other studies of the SDXB show that the EM of the Galactic halo varies by an order of magnitude in different sightlines \citep{Henley2013, Henley2015, Yoshino2009, Nakashima2018}. This large variation critically affects estimates of the path-length of the warm-hot gas and subsequently the mass measurements. Therefore it is essential to measure X-ray emission of the warm-hot gas close to the absorption sightline \citep{Gupta2014, Gupta2017}.

With the goal to determine the X-ray emission properties of the Milky Way CGM near absorption sightlines of \mrkn, \arkn, \ngcn, and \h (from \citet{Gupta2012}), we observed with \suzaku four empty fields close to these sightlines for $\rm 80~ks$ each (PI: Gupta). We also observed three of the \suzaku fields with \chandra to identify the points sources contaminating the diffuse X-ray emission (PI: Gupta). In this paper we report on the analysis of our  \chandra (\S2) and \suzaku (\S3) observations. In \S4 we present our results from the spectral analysis; we show that in addition to the gas close to the virial temperature, a super-virial temperature component or an enhanced Ne abundance is required. Our Discussion is presented in \S5, and we conclude in \S6. 

\section{Chandra Observations and Point Sources Detection} 
Even in ``empty" fields, point sources add a significant contribution to the truly diffuse emission, hence it is necessary to exclude them. Unfortunately, the poor angular resolution of \suzaku makes any significant point source identification impossible. We observed the \suzaku fields \offiiin, \offivn, and \offv with \chandra for $\rm \sim 10~ks$ each to identify the points sources. The observation IDs, dates, and exposure times are summarized in Table 1. The \offii close to the \mrk sight line had been observed with \xmm in 2014; for this field we identified the point sources from the \xmm observation. Detailed analysis of the \xmm \offii observation is described in \citet{Gupta2017}.

All the \chandra analysis was performed with the software \chandra interactive analysis of observations (CIAO 4.12)\footnote{https://cxc.harvard.edu/ciao/}. We reprocessed the \chandra data with the {\it repro} script to apply standard corrections. We also filtered the data sets for bad time intervals affected by flares. We generated $\rm 0.5-2.0~keV$ images and the corresponding exposure maps. We used the {\it wavdetect} tool to identify the point sources in the images. We identified 14, 13, and 9 X-ray point sources in the \offiiin, \offivn, and \offv images, respectively, down to the detection limit of $\rm 5.3 \times 10^{-15}~erg~cm^{-2}~s^{-1}$. In the \suzaku field \offii we had identified 23 point sources using the \xmm observation, with the detection limit of $\rm 2.6 \times 10^{-15}~erg~cm^{-2}~s^{-1}$. Our number of identified point sources are well in agreement with the Log$N$-Log$S$ in the Chandra Deep Field South (CDFS; \citet{Lehmer2012}). 

\section{Suzaku Observations and Data Reduction} 
In this work we only used data from the back-illuminated X-ray Imaging Spectrometer (XIS-1), featuring the largest effective area among all the \suzaku detectors at soft X-ray energies ($\rm 0.3-5.0~$ keV). The \suzaku observation IDs, dates, pointing directions, and exposure times are summarized in Table 1.

For data reduction we followed the procedures as described in the \suzaku Data Reduction Guide\footnote{https://heasarc.gsfc.nasa.gov/docs/suzaku/analysis/abc/}. In addition to standard data processing, we performed  data screening with the cut-off-rigidity (COR) of the Earth’s magnetic field, which varies as \suzaku traverses its orbit. During times with larger COR values, fewer particles are able to penetrate the satellite and the XIS detectors. We excluded times when the COR was less than $8~$GV, which is greater than the default value of $2~$GV. This helps to lower the particle background. 

Our goal is to extract the spectrum of the diffuse emission. Therefore we first removed the point sources identified in \chandra and \xmm observation, and then extracted the spectrum from the entire field-of-view. Since \suzaku has a broad point spread function (half-power diameter $\rm \sim 2\arcmin$), we excised regions within $\rm 1\arcmin$ radius of point source locations from the \suzaku fields. For sources too bright to be adequately removed by this region, we selected larger source exclusion radii by eye of the order of $\rm 2\arcmin - 3\arcmin$.

We produced the redistribution matrix files (RMFs) using the {\it xisrmfgen} ftool, in which the degradation of energy resolution and its position dependence are included. We also prepared ancillary response files (ARFs) using {\it xissimarfgen} ftool with the revised recipe\footnote{ https://heasarc.gsfc.nasa.gov/docs/suzaku/analysis/xisnxbnew.html}. For the ARF calculations we assumed a uniform source of radius $\rm 20\arcmin~$ and used a detector mask which removed the bad pixel regions. We estimate the total instrumental background from the database of the night Earth data with the {\it xisnxbgen} ftool. 

\section{Spectral Analysis}

\subsection{Milky Way CGM Emission}
The goal of this study was to constrain the properties of the Milky Way CGM. We isolated the Milky Way CGM from the different components of the SDXB (empty field spectrum) by spectral analysis. This was a challenging task as SDXB has contributions from different sources such as the Solar Wind Charge eXchange (SWCX), the Local Bubble (LB), the cosmic X-ray background (CXB) made of unresolved point sources, and the Milky Way CGM itself. 

Therefore, a typical SDXB spectrum is described by a three components model: 1) a foreground component consists of LB and SWCX, modeled as an unabsorbed thermal plasma emission in collisional ionization equilibrium (CIE) with a temperature of $\rm kT = 0.1~ keV$ \citep{Liu2017, Henley2013, Gupta2009}, 2) a background component of CXB  modeled with an absorbed power-law, and finally 3) the Milky-Way CGM component, modeled as an equilibrium thermal plasma absorbed by the cold gas in the Galactic disk. We performed all the spectral fitting with Xspec version 12.10.1f~\footnote{https://heasarc.gsfc.nasa.gov/xanadu/xspec/}. We modeled all the thermal plasma components in CIE with the APEC (version 3.0.9) model \citep{Smith2001} and used solar relative metal abundances of \cite{Anders1989}. For absorption by the Galactic disk, we used the {\it phabs} model in XSPEC.

Initially we fit the \suzaku XIS-1 spectra with the standard SDXB three component model [$apec_{(LB+SWCX)} + phabs \times (apec_{Halo}+powerlaw_{CXB})$]. The temperature of the foreground component was frozen at $\rm kT=0.1~keV$, but we allowed the normalization to vary. We fixed the total metallicity to 1 (in solar units) for both the thermal components as there is a strong correlation between the total metallicity and normalizations (or EM). The Galactic column densities were fixed to values determined from \citet{Dickey1990}. We fixed the power-law photon index to 1.52 and left the normalization as free parameter in the spectral fits (Table 3). After fitting the spectra (in the energy range of $\rm 0.3-5.0~keV$) with the three component model as described above, we noted a significant excess in the data around $\rm 0.50-0.55~keV$ in all the four spectra (Figure 1). We also noted excess emission near $\rm 0.8-1.0~keV$ in \offiin, \offiii and \offv (Figure 2). We attempted to fit these excess emissions as discussed below.

%%\subsection{Excess Emission near 0.50-0.55 keV}
\subsubsection{OI Contamination in Suzaku Data}
\citet{Sekiya2014} noted that \suzaku observations, particularly after 2011, were affected by the increase in the solar activity. The enhanced interaction of solar X-rays with the neutral Oxygen in the  Earth's atmosphere creates a \oi fluorescent line at $\rm E = 0.525~keV$ and contaminates \suzaku spectra. The \suzaku XIS cannot distinguish the \oi line from the \ovii K$\alpha$ triplet ($561, 569$ and $574$ eV, centroid at $571$ eV ) owing to its energy resolution of $\rm \approx 50~ eV$. Unless the \oi fluorescent line is taken into account in the spectral analysis, the \ovii line intensity would be overestimated. 

Since all the observations in this study were taken in 2014, the excess near $\rm 0.50-0.55~keV$ could be due to the \oi fluorescent line. To investigate the effect of the \oi contamination we measured the intensity of oxygen emission lines. We modified the above three-component model by switching the APEC thermal plasma component to VAPEC, which allows for variable elemental abundances. We fixed the oxygen abundance to zero, thereby removing oxygen contribution from the model. We then added three Gaussian lines at the energies of the \oi (525 eV), \ovii K-$\alpha~$triplet (centroid at 571 eV), and \oviii (665 eV) lines, and fitted the spectra again. Table 2 reports the best-fit oxygen line intensities. Note that  \oi has significant contribution in all spectra ranging from $2.3$ to $\rm 20.7 ~photons ~s^{-1}~cm^{-2}~str^{-1}~ (L.U.)$. This is about $25$\% to $130$\% of the \ovii intensity in the four spectra.  

The \oi contamination can be minimized by choosing the events taken during time intervals when the elevation angle from the bright Earth limb (DYE\_ELV parameter) is greater than $\rm 60^{\circ}$ \citep{Sekiya2014}. However, this results in the loss of a large amount of data. To further investigate the \oi contamination, we compared the \oi line intensity for DYE\_ELV $>20^{\circ}$, $>40^{\circ}$ and $>60^{\circ}$ in our observations (Table 2). \offiv had unusually high \oi emission ($\rm 20.75\pm1.38~ L.U.$) for the DYE\_ELV $>20^{\circ}$ selection, but with DYE\_ELV $>40^{\circ}$ \oi emission reduced significantly. With DYE\_ELV $>40^{\circ}$, \oi emission was reduced to $6$\% to $40$\% of the \ovii intensity. Therefore instead of restricting the data filtered only for DYE\_ELV$>60^{\circ}$, for further analysis we used DYE\_ELV $>40^{\circ}$; we took care of the residual \oi contamination by adding a Gaussian line to our spectral model. This resulted in a good balance between optimizing the effective exposure time and mitigating the \oi contamination. As discussed in the next section, the use of DYE\_ELV $>40^{\circ}$ instead of the more stringent limit of DYE\_ELV $>60^{\circ}$ has not caused any bias in our measurements.

\subsubsection{One-Temperature 1T CGM Model}
We refitted the \suzaku XIS-1 spectra with the standard SDXB three component model plus a Gaussian line to account for the \oi emission. We call this model the one-temperature (1T) CGM model. Figure 1 shows the best fit 1T model for \offiin. 

As noted above, we used data with DYE\_ELV$>40^{\circ}$ to optimize the exposure time, and so obtain tighter constraints on the fit statistics. To further verify that this choice is not introducing any bias in our analysis, we also fitted the 1T model to data sets filtered with DYE\_ELV $>20^{\circ}$ and $>60^{\circ}$. Table 3 reports the best fit 1T model parameters for DYE\_ELV $>20^{\circ}$, $>40^{\circ}$, and $>60^{\circ}$ for our observations. Best fit parameters are consistent within uncertainties and DYE\_ELV $>40^{\circ}$ provides the better fit statistics,  especially for \offiv and \offvn.

The temperatures of the Galactic CGM (absorbed thermal component) among all the fields are consistent with each other within the uncertainties, with an average value of $\rm kT = 0.195\pm 0.007 ~keV$, close to the Galaxy's virial temperature. The EMs vary between $\rm 5.7 - 14.3 \times 10^{-3}~cm^{-6}~pc$ range (Table 3); these EMs are very high, well above the typical values reported by previous studies (we discuss this in detail below in \S5).

In our 1T spectral model the CXB from unresolved extragalactic sources is modeled with an absorbed power-law with a fixed photon index of $\rm 1.52$. The best fit normalizations at $\rm 1~keV$ are in the range of $\rm 8.7-13.2~ photons~keV^{-1}~s^{-1}~sr^{-1}~cm^{-2}$ (Table 3). \citet{Cappelluti2017} using \chandra deep observations of the COSMOS field provide one of the most accurate measurements of the CXB. They have measured the $\rm 1 ~keV$ normalization of the CXB power-law of $\rm 10.91\pm0.16~ photons~keV^{-1}~s^{-1}~sr^{-1}~cm^{-2}$. The power-law normalizations in our observations are in the same range as the COSMOS field and other previous \suzaku and \xmm studies of SDXB \citep{Galeazzi2007, Gupta2009, Nakashima2018}. Thus our results are not biased by the the fit to the CXB component.

We modeled the foreground emission of LB and SWCX as an unabsorbed plasma in CIE with thermal emission. We fixed the temperature of this component to $\rm kT = 0.1~ keV$ and measured the best fit EMs in the range of $\rm 0.015-0.018~cm^{-6}~pc$ toward our four fields (Table 3). To determine what fraction of the foreground emission is due to SWCX, which is a time variable component, we first estimated the LB emission using the maps in \citet{Liu2017}. They have generated all sky maps of the LB temperature and EM using the ROSAT All Sky Survey (RASS) data corrected for the SWCX. In Liu et al. maps the LB EMs are near $\rm \sim 0.001~cm^{-6}~pc$ toward \offiin, \offiiin, \offv and $\rm \sim 0.003~cm^{-6}~pc$ towards \offivn. The LB EMs are much lower than our measured values of foreground EMs of $\rm 0.015-0.018~cm^{-6}~pc$, clearly showing high contribution from the SWCX. However, this is not unexpected as our observations were made during 2014, which was a period of high solar activity. Some of the previous studies of SDXB have also measured such large EMs for the foreground component \citep{Nakashima2018, Henley2015, Gupta2009}. \citet{Henley2015} well-constrained the foreground component EMs in the range of $\rm 0.005-0.079$ (for \citet{Anders1989} abundances) using the \suzaku and \xmm shadow observations, bracketing our values. Thus our results do not seem to be biased by the fit to the foreground components either.

We used our fit results of the foreground (unabsorbed APEC) component to obtain \ovii and \oviii line intensities from the LB and the SWCX. We measured total \ovii line intensities from LB plus SWCX of $\rm 5.89\pm1.55$, $\rm 5.20\pm0.97$, $\rm 6.05\pm1.20$ and $\rm 5.44\pm1.43$ LU toward \offiin, \offiiin, \offiv, and \offvn, respectively. The \oviii line intensities are $\rm 0.30\pm0.08$, $\rm 0.26\pm0.05$, $\rm 0.30\pm0.06$, \& $\rm 0.27\pm0.07$ LU for \offiin, \offiiin, \offiv, and \offvn, respectively. The LB EMs from Liu et al. of $\rm 0.001-0.003$ corresponds to \ovii and \oviii line intensities of $\rm 0.33-1.00~LU$ and $\rm 0.02-0.06~LU$. Subtracting LB emission from the total foreground line intensities can provide estimate of contribution from the SWCX \ovii and \oviii line emission. The estimated SWCX \ovii and \oviii line intensities in our observations are in the range of $\rm \sim 4.9-5.6$ LU and $\rm \sim 0.24-0.28$ LU.

%%\subsection{Excess Emission near 0.8-1.0 keV}
\subsubsection{Two-Temperatures (2T) or Enhanced Ne Abundance (1T Ne) CGM Models}
After fitting the spectra with the 1T models, we found excess emission in the $\rm 0.8-1.0~ keV$ range   along three out of four sight lines (Figure 2). We attempted to model this excess emission in two different ways. First we added another absorbed thermal component (APEC) to the 1T model; we refer to this as the two temperature (2T) model. Thus the 2T model is given by $apec_{LB+SWCX} + phabs \times (apec_{Halo}+apec_{Halo}+powerlaw_{CXB}) + gaussian_{OI}$. Model fits to these three spectra were  significantly better with the 2T model compared to the 1T model ($\rm \Delta \chi^{2}/\Delta d.o.f.=10.5/2,~15.3/2,~8.9/2$ for the \offiin, \offiiin, and \offv spectra, respectively). The additional thermal component is required at a significance of 99.1\%, 99.9\%, and 97.6\% (F-test) for the \offiin, \offiiin, and \offv spectra, respectively. The temperature of the warm-hot CGM is consistent within the errors with an average value of $\rm kT = 0.177\pm 0.009 ~keV$ and EM is between $\rm 7.2 - 17.4 \times 10^{-3}~cm^{-6}~pc$ (Table 4). The temperature of the second  component $\rm (kT = 0.6-0.9~ keV)$ is much higher than the warm-hot phase; we call this the hot component here onward (Figure 2 and Table 4).

The hot component spectrum is peaked at around $\rm \approx 0.9~keV$, close to the energy of the \neix forbidden line ($\rm 0.92~ KeV$). Along with  \ovii and \oviiin,  \neix also probes the warm-hot ($\rm \sim 10^{6}~K$) medium. Therefore we tried to fit the spectra with the 1T model, but replacing the APEC model with VAPEC, with variable Ne abundance (we call this the $\rm 1T_{Ne}$ model); abundances of other elements were fixed at 1 solar as before (as done in \citet{Mitsuishi2012, Yoshino2009}). This model also showed a similar improvement in the fit statistics ($\rm \Delta \chi^{2}/\Delta d.o.f.=9.4/1, ~15.6/1, ~10.6/1$ for the \offiin, \offiiin, and \offv spectra, respectively; Table 4). The Ne overabundance is required at significance of  99.6\%, 99.9\%, and 99.8\% (F-test) for the \offiin, \offiiin, and \offv spectra, respectively. The temperature of the warm-hot CGM in the three fields is similar within the uncertainties with an average value of $\rm kT = 0.190\pm 0.007 ~keV$, EM is between $\rm 6.0 - 14.6 \times 10^{-3}~cm^{-6}~pc$ and the Ne abundances are in the range of $\rm 1.4-3.8$ times solar (Table 4). 

\subsubsection{Simultaneous Fit}
The temperature of the warm-hot component was found to be consistent within errors among all fields ($\rm kT_{1} \approx 0.2 ~keV$; Table 4). Therefore we performed global fits, simultaneously fitting all the four spectra for 1T, 2T, and $\rm 1T_{Ne}$ models. We tied the temperature of the warm-hot component and allowed the normalization to vary among different fields. We also allowed to vary all other free model parameters among observations.

The resulting warm-hot phase temperature was $\rm 0.198\pm0.007~keV$ for 1T and $\rm 1T_{Ne}$ and $\rm 0.176\pm0.008~keV$ for 2T models. The warm-hot phase EMs are in the range of $\rm 0.7-1.8 \times 10^{-2}~ cm^{-6}~pc$. The hot component has temperatures and EMs in the range of $\rm 0.65-0.90~keV$ and $\rm 0.4-1.0 \times 10^{-3}~ cm^{-6}~pc$, respectively. In the $\rm 1T_{Ne}$ model, the Ne abundances are in the range of $\rm 1.4-4.0$ times solar. The hot component or the overabundance of Ne are required at a significance of more than $4\sigma$ (F-test probability of $\rm > 99.99$\%; $\rm 2T~ model: \Delta \chi^{2}=61.6,~\Delta d.o.f.=8;~1T_{Ne}~ model: \Delta \chi^{2}=44.6,~\Delta d.o.f.=4$). The simultaneous fit parameters are reported in Table 5. 

\subsection{Point Sources Contribution}
To estimate the contribution of point sources to empty field observations, we have extracted the combined spectrum of all sources detected in each \suzaku field-of-view. For the background spectrum we used the diffuse emission spectrum extracted from the entire field after removing the detected sources. The background was scaled by the ratio of the total area of the sources to the area of diffuse background. The point sources occupy about $\rm 30-40\%$ area of the entire field-of-view in our observations.

We fitted the cumulative point sources spectrum for each field with an absorbed power-law. The best fit powerlaw photon-indices varies from 1.98 to 2.69 in our fields. The spectrum of the sources identified in the \offiii shows an excess emission over powerlaw around $\rm ~1~ keV$. The excess is fitted well with a thermal component with temperature $\rm T = 1.02~keV$ and $\rm 0.5-2.0~keV$ flux of $\rm 1.8 \times 10^{-13}~erg~s^{-1}~cm^{-2}$. This excess thermal component could be due to Milky Way stellar sources. Many stars are known to have two temperature thermal spectra, a hot active component at a nominal temperature of $\rm kT_{a} \sim 1 ~keV$ and a comparatively cooler quiescent component at $\rm kT_{q} \sim 0.3~keV$ \citep{Kashyap1992}. \cite{Gupta2009b} studied the point sources X-ray emission identified in the high latitude \xmm fields. The authors also noted similar thermal component emission with a temperature $\rm T = 0.92 ~keV$ and they attributed it to the stellar hot active component. 

In our fields the total $\rm 0.5-2.0~keV$ surface brightness (SB) of X-ray background (diffuse+resolved point sources) are $\rm 2.2 \times 10^{-11}, ~1.5 \times 10^{-11}, ~ 1.6 \times 10^{-11}$, and $\rm 1.6 \times 10^{-11}~ergs~s^{-1}~cm^{-2}~deg^{-2}$ along \offiin, \offiiin, \offivn, and \offvn, respectively. We measured the point sources SB of $\rm 3.3 \times 10^{-12},~2.0 \times 10^{-12}, ~ 2.2 \times 10^{-12} $, and $\rm 9.7 \times 10^{-13}~ergs~s^{-1}~cm^{-2}~deg^{-2}$, which corresponds to $\rm 6-15\%$ of the total SB of an empty field emission. Thus it is unlikely that the details of the point-source subtraction process has biased our results in any way.

\section{Discussion}
\subsection{Hot or Ne Overabundant Phase of the CGM}
In three out of our four \suzaku observations we have detected excess soft X-ray emission near $\rm 0.8-1.0~keV$. We used two different models to fit this excess emission. We found that this excess emission is either from the hot gas at temperatures near $\rm 0.65-0.90~keV$ with EM of $\rm 1.0\pm0.4~\times 10^{-3}~ cm^{-6}~pc$ or from an enhanced Ne abundance of $\rm 1.5-4.0$ solar in the warm-hot gas in the Galactic CGM. The temperature of the hot phase is similar to the recently discovered hot component in the Milky Way CGM by \citet{Das2019a, Das2019c}. Using a very high S/N \xmm absorption and emission spectra in the sightline of blazar $1ES~1553+113$, Das et al. discovered the hot $\sim 0.86~$keV gas coexisting with the warm-hot $\sim 0.09~$keV gas in the Galactic CGM. This was a robust detection and it was the first time that the hot component was detected both in emission and absorption. The $1ES~1553+113$ sightline passes close to the X-ray shell around the Fermi Bubble (FB); while Das et al. concluded that in absorption the hot gas is unlikely to be associated with the FB, this possibility could not be ruled out for the hot phase detected in emission. Das et al. also noted non-solar abundance ratios and alpha-enhancement of light metals, N, O, Ne, in the warm-hot phase, similar to our Ne overabundance.

A few earlier \suzaku and \xmm X-ray emission studies also reported suggestive evidence of a higher temperature ($\rm 0.5-0.9~keV$) or enhanced Ne abundance in the Galactic halo. \citet{Ursino2016} reported an excess emission around $\rm 0.9~keV$ towards the inner region of the North Polar Spur (NPS)/Loop1 structure. They probed the region using \suzaku shadow observations of the high column density cloud {\it MBM36} and a nearby empty region. The authors modeled this excess emission with the Ne overabundance [$\rm Ne/O=1.7$] or an absorbed hotter thermal component at $\rm kT=0.76~keV$ with EM of $\rm 1.7\pm0.6 \time 10^{-3}~ cm^{-6}~pc$. Since this sight line probes the NPS/Loop 1 structure, they associated this excess emission to the shell of a superbubble predicted by some NPS models. Our measured hotter halo component temperatures/EMs or enhanced Ne abundances are in the same range as observed by \citet{Ursino2016}. 

\citet{Yoshino2009} analyzed the soft diffuse X-ray emission of 13 high latitude \suzaku fields. For their preferred model, they used variable Fe and Ne abundances for the Galactic halo thermal emission. They measured the average temperature of $\rm 0.216\pm0.017~keV$ and Fe/O and Ne/O ratios of $0.51-2.99$ and $0.91-3.79 $, respectively. They fitted spectra toward four sightlines with supersolar Fe and Ne abundances with an alternative model of additional higher temperature ($\rm 0.5-0.9 ~keV$) emission component with solar abundances (similar to our 2T model). Their measured EMs of higher temperature component of $\rm 5.0-9.5~\times 10^{-4}~ cm^{-6}~pc$ or Ne/O ratios of $0.91-3.79 $ are in the same range as our values. 

\citet{Henley2013} also reported detection of a high halo temperature ($\rm T \approx 0.86~$keV) towards one sight line ($\# 83$, their table 1) and some excess emission around $\rm \approx 1 ~keV$ in some other sightlines. For the $\# 83$ sightline they used a 2T model, one thermal component to model the excess emission around $\rm \approx 1 ~keV$, and one to model the $\rm 0.17-0.26~keV$ emission. The authors did not report on such excess emission in other sight lines. 

In Figure 3, we have plotted all these observations on the sky map, which shows a widespread prevalence of the hot component of the CGM and/or Ne overabundance.  Our sightlines are also far from the NPS/Loop1 structure, or from the Fermi bubble, suggesting that the presence of the hot component or non-solar enhanced Ne gas are not necessarily associated with these special structures in the Galactic center. 

\subsection{Very Bright Warm-Hot Phase of the CGM}
The average temperature of the warm-hot phase of the CGM that we measure, $\rm \approx 0.2~keV$, is close to the Milky Way virial temperature and is in agreement with the previous studies of the Galactic halo X-ray emission. However, our measured EMs are excessively high ($\rm 1.13-1.74~\times 10^{-2}~ cm^{-6}~pc$) along \offiin, \offivn, and \offvn. \offii was also observed with \xmm in 2014 and we had found a similarly high EM of $\rm \sim 1.65 \times 10^{-2}~cm^{-6}~pc$ \citep{Gupta2017}.

\cite{Henley2015} measured Galactic halo EMs in the range of $\rm 2.2-6.7 \times 10^{-3} cm^{-6}~{pc}$\footnote{for their LB foreground model and the Anders \& Grevesse (1989) abundances; same as  our 1T model.} using the \xmm and/or \suzaku shadow observations toward six sight-lines. \cite{Henley2013} studied the halo emission along $110$ high-latitude \xmm sight-lines and found the EM varies by over an order of magnitude ($\rm 0.4-7 \times 10^{-3}~cm^{-6}~pc$ with median detection of $\rm 1.9 \times 10^{-3}~cm^{-6}~pc$). Our measure CGM EMs toward three fields are significantly higher than these values.

Only recently using the HaloSat data \citet{Kaaret2020} found similarly high values of the warm-hot CGM EMs $\rm 1.1-1.4 \times 10^{-2}~cm^{-6}~pc$ (converting their EMs to EMs for solar abundances) toward the inner halo. Kaaret et al. measured Galatic halo temperature and EM of $\rm 0.20~keV$ and $\rm 1.26 \times 10^{-2}~cm^{-6}~pc$, respectively, in one of their fields close to our \offv sightline. This suggests that the excessively high EMs are associated with the inner halo. In December 2020, when we were revising this paper after the referee comments, a newly published sky map from the first eROSITA all-sky survey reported the discovery of large ($\rm \sim 14~ kpc$ above and below the Galactic centre) soft-X-ray-emitting bubbles \citep{Predehl2020}. The high EMs toward our fields could also be a result of the contribution from these bubbles. Our sight lines of \offii and \offv pass through the southern bubble and \offiv sightline pass close to the outer boundary of the northern bubble. We will investigate this further in a subsequent paper (Gupta et al., in preparation).

\section{Conclusion}
In this work we present the soft X-ray emission properties of the Galactic CGM along four directions using \suzaku and \chandra observations. 
We have found strong \oi contamination in the \suzaku data using standard filtering for the elevation angle from the bright Earth limb of larger than $\rm DYE\_ELV>20^{\circ}$. We carefully investigated the effect of \oi contamination and found with $\rm DYE\_ELV>40^{\circ}$, \oi emission can be reduced significantly, and a good balance is struck between the loss of exposure time and \oi contamination. To model the residual \oi emission we included a Gaussian emission  at $\rm 0.525~keV$ in the spectral modeling. 

We clearly detect the emission from the warm-hot CGM of the Milky Way in all the four sightlines. The measured temperature of $\rm \approx 0.2~keV$ ($\rm 2.1 \times 10^{6}~K$) is close to the Galaxy's virial temperature, consistent with previous studies. However, the EMs are very high toward three sightlines probing the inner halo. These high EMs are in line with the recent HaloSat results \citep{Kaaret2020}.

Toward three sightlines, we have detected excess emission near $\rm 0.8-1.0~keV$. There are two possibilities to explain this excess emission. It may arise from a hotter component of the Galactic CGM  at a temperature of about $\rm 0.65-0.90~KeV$. The temperature of this hotter component is similar to that recently discovered ($\sim 0.86~$KeV) by \citet{Das2019a, Das2019c} towards the sightline to blazar $1ES~1553+113$. Detection of X-ray emission from the hot CGM in our \suzaku fields suggest that the $1ES~1553+113$ sightline is not unique and the hot phase of the Galactic CGM is more widespread. Alternatively, the excess emission could be a signature of overabundance of Ne in the warm-hot phase of the Galactic CGM. The variable Ne abundance model required Ne/O abundances of $1.5-3.4$ solar; this is also not unprecedented (\S 5.1). Both the models of the excess emission have similar statistical significance, so we do not prefer one over the other. Future missions like XRISM and Athena with better spectral resolution may able to differentiate the two models.

Whether the excess emission is from the hot gas or overabundance Ne, it poses a challenge to galaxy formation and evolution models. Literature on super-virial temperature and/or non-solar abundance ratios in the CGM is limited. A multi-stage feedback can produce the hot component as suggested by the theoretical models of \cite{Tang2009}. As discussed in Das et al., the non-solar Ne/O abundance ratio may be informative of the depletion of metals like oxygen. Recently, \cite{Voort2020} have studied the effect of magnetic fields on a simulated galaxy and its CGM. Their results show that magnetic fields in the galaxy's halo can cause inhomogeneous mixing; this might lead to apparent non-solar abundance ratios. Understanding the origin of the super-virial hot component and/or non-solar abundance ratios of the Milky Way CGM is going to be of interest for theoretical models of the CGM. 

\section{Acknowledgements}
We thank the anonymous referee for the useful comments that helped improve the paper. We gratefully acknowledge support through the NASA ADAP grant $\rm 80NSSC18K0419$ to AG. Support  for  this  work  was  also  provided  by the  National  Aeronautics  and  Space  Administration through Chandra Award Number $\rm GO9-20077X$ to AG issued by the Chandra X-ray Observatory Center, which is operated by the Smithsonian Astrophysical Observatory for and on behalf of the National Aeronautics Space Administration under contract NAS8-03060. SM gratefully acknowledges the NASA grant NNX16AF49G. YK acknowledges support from PAPIIT grant IN106518 and PAPIIT-PASPA.

%\newpage

%%%%%%%%%%%%%%%%%%%%%%%%%%%%%%%%%%%%%%%%%%%%%%%%%%%%%%%%%%%%%%%%%%%%%%%%%%

\begin{deluxetable*}{lccccccccc}
\tabletypesize{\scriptsize}
\tablewidth{0pt} 
\tablenum{1}
\tablecaption{Suzaku and Chandra Observations Log \label{tab:deluxesplit}}
\tablehead{
\colhead{Target} & \colhead{\it l}& \colhead{\it b} & \multicolumn{3}{c}{Suzaku} & \colhead{     } & \multicolumn{3}{c}{Chandra}\\
\cline{4-6}
\cline{8-10}
\colhead{} & \colhead{} & \colhead{} & \colhead{ObsID}& \colhead{Start Date} & \colhead{Exposure} & \colhead{    }& \colhead{ObsID}& \colhead{Start Date} & \colhead{Exposure}\\
\colhead{} & \colhead{($^{\circ}$)}& \colhead{($ ^{\circ}$)} & \colhead{}& \colhead{} & \colhead{(ks)} & \colhead{    }& \colhead{}& \colhead{} & \colhead{(ks)}
} 
%%\colnumbers
\startdata 
{Off-field 2} & $37.42$  & $-30.55$ & 509043010 & {May 7 2014}   & $80.16$ && \nodata & \nodata & \nodata \\
{Off-field 3} & $91.73$  & $-24.10$ & 509044010 & {June 11 2014} & $97.33$ && $21380$   & {July 14 2019}  &  $9.45$  \\
{Off-field 4} & $286.33$ & $+23.55$ & 509045010 & {Dec 17 2014}  & $81.32$ && $21381$   & {Nov 29 2018}  & $9.94$ \\
{Off-field 5} & $39.92$  & $-36.21$ & 509046010 & {May 10 2014}  & $80.91$ && $21382$   & {Nov 8 2018}  & $9.95$ \\
\enddata
\end{deluxetable*}
%%%%%%%%%%%%%%%%%%%%%%%%%%%%%%%%%%%%%%%%%%%%%%%%%%%%%%%%%%%%%%%%%%%%%%%%%%%%%%%%%%%%%%%%%%%%%%%%%%%

\begin{deluxetable*}{lccccc}
\tabletypesize{\scriptsize}
\tablenum{2}
\tablecaption{Oxygen Emission Line Intensities. }
\tablehead{\colhead{Data set} & \colhead{Exposure} & \colhead{$\rm O_{VII}$} & \colhead{$\rm O_{VIII}$} & \colhead{$\rm O_{I}$} & \colhead{$\chi^{2}/d.o.f.$}\\
\colhead{Data set} & \colhead{(sec)} & \colhead{L.U.} & \colhead{LU} & \colhead{LU} & \colhead{} \\
} 
%%\colnumbers
\startdata
{Off-field 2} &    &    &    &     &     \\
\hline
{DYE\_ELV $>20^{\circ}$  }  & $53580 $ &  $16.79\pm2.16$  &  $5.70\pm0.95$  & $5.86\pm1.42$    &   $195.3/202 $   \\
{DYE\_ELV $>40^{\circ}$  }  & $47609 $ &  $17.08\pm2.25$  &  $5.59\pm0.96$  & $3.12\pm1.68$    &   $178.8/177$  \\
{DYE\_ELV $>60^{\circ}$  }  & $44147 $ &  $17.55\pm2.35$  &  $5.67\pm1.02$  & $2.31\pm1.75$    &   $162.7/164 $  \\
\hline
{Off-field 3} &    &    &    &     &     \\
\hline
{DYE\_ELV $>20^{\circ}$  }  &  $79816 $ &  $9.31\pm1.29$  &  $2.25\pm0.58$  & $2.34\pm0.89$     &   $321.3/329 $   \\
{DYE\_ELV $>40^{\circ}$  }  &  $72709 $ &  $8.59\pm1.30$  &  $2.33\pm0.57$  & $1.37\pm1.04$     &   $310.6/301 $  \\
{DYE\_ELV $>60^{\circ}$  }  &  $63240 $ &  $8.58\pm1.37$  &  $2.45\pm0.62$  & $0.68\pm0.85$    &   $277.6/269 $  \\
\hline
{Off-field 4} &    &    &    &     &     \\
\hline
{DYE\_ELV $>20^{\circ}$  }  &  $62975 $ &  $16.04\pm1.63$  &  $2.58\pm0.63$  & $20.75\pm1.38$    &   $378.0/328 $   \\
{DYE\_ELV $>40^{\circ}$  }  &  $44121 $ &  $10.75\pm1.66$  &  $3.49\pm0.70$  & $4.25\pm1.21$     &   $252.0/238 $ \\
{DYE\_ELV $>60^{\circ}$  }  &  $30026 $ &  $10.19\pm1.85$  &  $3.20\pm0.83$  & $0.52\pm0.85$     &   $169.0/153 $  \\
\hline
{Off-field 5} &    &    &    &     &     \\
\hline
{DYE\_ELV $>20^{\circ}$  }  &  $52949 $ &  $15.94\pm1.85$  &  $4.18\pm0.81$  & $4.32\pm1.10$    &  $268.8/245 $    \\
{DYE\_ELV $>40^{\circ}$  }  &  $45449 $ &  $13.96\pm1.85$  &  $4.17\pm0.83$  & $0.89\pm1.05$    &  $219.9/211 $   \\
{DYE\_ELV $>60^{\circ}$  }  &  $39274 $ &  $13.77\pm1.98$  &  $4.43\pm0.90$  & $0.75\pm1.00$    &  $226.2/183$   \\
\enddata
\tablecomments{\\
Normalization of Gaussian lines are in units of L.U. ($\rm photons~s^{-1}~cm^{-2}~str^{-1}$).
}
\end{deluxetable*}

%%%%%%%%%%%%%%%%%%%%%%%%%%%%%%%%%%%%%%%%%%%%%%%%%%%%%%%%%%%%%%%%%%%%%%%%%%%%%%%%%%%%%%%%%%%%%%%%%%%%

\begin{deluxetable*}{lcccccccc}
\tabletypesize{\scriptsize}
\tablewidth{0pt} 
\tablenum{3}
\tablecaption{Best Fit Parameters: One Temperature (1T) Model.  \label{tab:deluxesplit}}
\tablehead{
\colhead{Target} & \colhead{$\rm O_{I}$} & \colhead{Foreground}& \colhead{$\rm N_{H}$ $^{c}$} & \multicolumn{2}{c}{Galactic Halo} & \colhead{PowerLaw} &  \colhead{$\chi^{2}/d.o.f.$}\\
\cline{5-6}
\colhead{Data set} & \colhead{$\rm Norm^{a}$ } &\colhead{EM$^{b}$}& \colhead{} & \colhead{kT} & \colhead{EM}&  \colhead{Norm$^{d}$}&  \colhead{}\\
\colhead{} & \colhead{} &\colhead{($\rm 10^{-2}cm^{-6}~pc$)}& \colhead{$10^{20}\rm cm^{-2}$} & \colhead{(keV)} & \colhead{($\rm 10^{-2}cm^{-6}pc$)} & \colhead{} & \colhead{} 
} 
\startdata 
\hline
{Off-field~2} &    &    &    &     &    & & \\
\hline
{DYE\_ELV $>20^{\circ}$  } & $5.8\pm1.4$  &  $1.84\pm0.44$  & $4.78$  & $0.200\pm0.012$ & $1.41\pm$0.23  &  $8.84\pm0.54$  & $212.9/205$ \\
{DYE\_ELV $>40^{\circ}$  } & $3.2\pm1.4$  &  $1.76\pm0.46$  & $"$     & $0.199\pm0.011$ & $1.43\pm0.25$  &  $8.84\pm0.58$  & $194.5/179$ \\
{DYE\_ELV $>60^{\circ}$  } & $2.7\pm1.5$  &  $1.80\pm0.48$  & $"$     & $0.199\pm0.011$ & $1.48\pm0.27$  &  $8.69\pm0.60$  & $181.6/167$ \\
\hline
{Off-field~3} &    &    &    &     &    & & \\
\hline
{DYE\_ELV $>20^{\circ}$} & $2.9\pm0.9$ &  $1.55\pm0.29$  & $3.76$ & $0.200\pm0.016$ & $0.63\pm0.16$  & $13.02\pm0.46$ & $345.6/332$ \\
{DYE\_ELV $>40^{\circ}$} & $1.5\pm0.9$ &  $1.55\pm0.29$  & $"$  & $0.207\pm0.018$  & $0.57\pm0.14$  & $13.11\pm0.49$  &  $324.5/304$ \\
{DYE\_ELV $>60^{\circ}$} & $0.6\pm0.7$ &  $1.60\pm0.32$  & $"$  & $0.204\pm0.019$ & $0.58\pm0.16$  & $13.21\pm0.52$  & $287.2/272$ \\
\hline
{Off-field~4} &    &    &    &     &    & & \\
\hline
{DYE\_ELV $>20^{\circ}$} & $18.2\pm1.7$ &  $1.67\pm0.42$  & $9.02$ & $0.158\pm0.014$ & $2.03\pm0.74$  & $13.21\pm0.46$ & $436.8/331$ \\
{DYE\_ELV $>40^{\circ}$} & $2.6\pm1.2$ &  $1.81\pm0.36$  & $"$  & $0.183\pm0.014$  & $1.13\pm0.32$  &   $12.64\pm0.56$  &  $253.4/241$ \\
{DYE\_ELV $>60^{\circ}$} & $0.0\pm0.0$ &  $1.66\pm0.39$  & $"$  & $0.191\pm0.014$ & $1.03\pm0.25$  &   $11.13\pm0.64$  & $173.6/156$ \\
\hline
{Off-field~5} &    &    &    &     &    & & \\
\hline
{DYE\_ELV $>20^{\circ}$} & $3.2\pm1.3$ &  $1.51\pm0.41$  & $5.24$ & $0.185\pm0.010$ & $1.39\pm0.29$ & $10.85\pm0.52$ & $293.3/248$ \\
{DYE\_ELV $>40^{\circ}$} & $0.1\pm0.4$ &  $1.63\pm0.43$  & $"$  & $0.192\pm0.011$  & $1.18\pm0.22$  & $10.85\pm0.57$  &  $232.6/214$ \\
{DYE\_ELV $>60^{\circ}$} & $0.0\pm0.0$ &  $1.44\pm0.44$  & $"$  & $0.191\pm0.010$ & $1.23\pm0.21$  &  $10.75\pm0.60$  & $233.5/186$ \\
\enddata
\tablecomments{\\
$^{a}$ Normalization of the Gaussian at the fixed center energy E=0.525 keV in units of L.U. ($\rm photons~s^{-1}~cm^{-2}~str^{-1}$).\\
$^{b}$ Emission measure for foreground (LB+SWCX) component with temperature fixed at kT=0.1 keV.\\
$^{c}$ Galactic values of the absorption column density given in Dickey \& Lockman (1990)\\
$^{d}$ Normalization of the power-law model with fixed photon index $\rm \Gamma=1.52$ in the units of $\rm photons~keV^{-1}~s^{-1}~sr^{-1}~cm^{-2}$.\\
}
\end{deluxetable*}

%%%%%%%%%%%%%%%%%%%%%%%%%%%%%%%%%%%%%%%%%%%%%%%%%%%%%%%%%%%%%%%%%%%%%%%%%%%%%%%%%%%%%%%%%%%%%%%%%%%

\begin{deluxetable*}{lccccccccc}
\tabletypesize{\scriptsize}
\tablewidth{0pt} 
\tablenum{4}
\tablecaption{1T, 2T and 1T$_{Ne}$ model fits: Individual Spectral Fits}
\tablehead{
\colhead{Target/} & \colhead{$\rm O_{I}$} & \colhead{Foreground} & \multicolumn{5}{c}{\bf Galactic Halo} & \colhead{PowerLaw} &\colhead{$\chi^{2}/d.o.f.$}\\
\cline{4-8}
\colhead{Model} & \colhead{Norm$^{a}$} & \colhead{EM$^{b}$} & \colhead{kT$_{1}^{c}$} & \colhead{EM$_{1}$} &\colhead{kT$_{2}^{d}$} & \colhead{EM$_{2}$} & \colhead{Ne$^{e}$} & \colhead{Norm$^{f}$}  & \colhead{}  \\
\colhead{} & \colhead{} & \colhead{($\rm 10^{-2}cm^{-6}pc$)} & \colhead{(keV)} & \colhead{($\rm 10^{-2}cm^{-6}pc$)} &\colhead{(keV)} & \colhead{($\rm 10^{-2}cm^{-6}pc$)} & \colhead{}  &\colhead{} & \colhead{} \\
} 
\rotate
\startdata 
{\bf Off-field~2} & & & & & & & & & \\
\hline
1T$^{I}$ & $3.2\pm1.4$    & $1.76\pm0.46$   & $0.199\pm0.011$    &  $1.43\pm0.25 $ & \nodata & \nodata & \nodata  &  $8.84\pm0.58$   &  $194.5/179 $  \\
2T$^{II}$ &  $2.1\pm1.7$   & $1.49\pm0.55 $   & $0.178\pm0.016  $   & $1.74\pm0.44 $   & $0.691\pm0.152 $   & $0.082\pm0.042 $  & \nodata &  $8.27\pm0.65$  & $184.0/177 $\\
1T$_{\scriptsize Ne}$$^{III}$ &  $3.0\pm1.5$   & $1.75\pm0.48 $    & $0.195\pm0.011 $   &  $1.46\pm0.27 $   & \nodata   &   \nodata  &  $1.9\pm0.5 $    &  $8.65\pm0.59$    & $185.1/178 $  \\
\hline
{\bf Off-field~3} & & & & & & & & &\\
\hline
1T$^{I}$ &  $1.5\pm0.9$   & $1.55\pm0.29 $    &  $0.207\pm0.018 $      & $0.57\pm0.14 $ & \nodata & \nodata & \nodata  &  $13.11\pm0.49$   & $324.5/304 $  \\
2T$^{II}$ &  $1.0\pm0.9$   & $1.43\pm0.32$   & $0.178\pm0.020 $   & $0.72\pm0.25 $   & $0.799\pm0.121 $   & $0.059\pm0.024 $   & \nodata &  $12.45\pm0.55$   & $309.2/302 $ \\
1T$_{\scriptsize Ne}$$^{III}$ &  $1.3\pm0.9$   & $1.53\pm0.30$   &  $0.197\pm0.017 $   &  $0.60\pm0.16 $     & \nodata & \nodata & $2.9\pm0.9 $   & $12.83\pm0.49$    & $308.9/303 $\\
\hline
{\bf Off-field~4} & & & & & & & & &\\
\hline
1T$^{I}$ &  $2.6\pm1.2$   & $1.81\pm0.36$ & $0.183\pm0.014$   & $1.13\pm0.32 $ & \nodata & \nodata & \nodata &  $12.64\pm0.56$   &    $253.4/241$ \\
2T$^{II}$ &  \nodata   & \nodata    & \nodata    & \nodata    & \nodata    & \nodata  & \nodata & \nodata  & \nodata \\
1T$_{\scriptsize Ne}^{III}$ &  $2.6\pm1.2$   & $1.81\pm0.40$     & $0.181\pm0.014$    &  $1.14\pm0.33$     & \nodata & \nodata & $1.3\pm0.7$ & $12.64\pm0.57$  & $252.9/240$ \\
\hline
{\bf Off-field~5} & & & & & & & & &\\
\hline
1T$^{I}$ &  $0.1\pm0.4$   &  $1.63\pm0.43 $ & $0.192\pm0.009 $   & $1.18\pm0.22 $ & \nodata & \nodata & \nodata & $10.85\pm0.57$    & $232.6/214$\\
2T$^{II}$ &  $0.0\pm0.0$   & $1.48\pm0.46 $   & $0.176\pm0.012 $   & $1.34\pm0.28 $   & $0.690\pm0.125 $   & $0.060\pm0.032 $ & \nodata &  $10.28\pm0.63$   & $223.7/212 $ \\
1T$_{\scriptsize Ne}$$^{III}$ &  $0.0\pm0.0$   & $ 1.61\pm0.43$     & $0.187\pm0.010 $    &  $1.22\pm0.22 $    & \nodata & \nodata & $2.2\pm0.7 $ & $10.66\pm0.56$  & $222.0/213$\\
\enddata
\tablecomments{\\
$^{a}$ Normalization of the Gaussian at the fixed center energy E=0.525 keV in units of L.U. ($\rm photons~s^{-1}~cm^{-2}~str^{-1}$).\\
$^{b}$ Emission measure for foreground (LB+SWCX) component with temperature fixed at kT=0.1 keV.\\
$^{c}$ Temperature of Galactic halo warm-hot phase thermal component.\\
$^{d}$ Temperature of Galactic halo hot phase thermal component.\\
$^{e}$ Ne abundance for Galactic halo warm-hot phase component.\\
$^{f}$ Normalization of the power-law model with fixed photon index $\rm \Gamma=1.52$ in the units of $\rm photons~keV^{-1}~s^{-1}~sr^{-1}~cm^{-2}$.\\
$^{I}$ Galactic halo one temperature model.\\
$^{II}$ Galactic halo two temperature model.\\
$^{III} $Galactic halo one temperature model with variable Ne abundance.
}
\end{deluxetable*}

%%%%%%%%%%%%%%%%%%%%%%%%%%%%%%%%%%%%%%%%%%%%%%%%%%%%%%%%%%%%%%%%%%%%%%%%%%%%%%%%%%%%%%%%%%%%%%%%%%%

\begin{deluxetable*}{lcccccccccc}
\tabletypesize{\scriptsize}
\tablewidth{0pt} 
\tablenum{5}
\tablecaption{1T, 2T and 1T$_{Ne}$ model fits: Simultaneous Spectral Fits}
\tablehead{
\colhead{Model/} & \colhead{$\rm O_{I}$} &\colhead{Foreground} & \multicolumn{5}{c}{\bf Galactic Halo} & \colhead{PowerLaw} &\colhead{$\chi^{2}/d.o.f.$}\\
\cline{4-8}
\colhead{Target} & \colhead{Norm$^{a}$} &\colhead{EM$^{b}$} & \colhead{kT$_{1}^{c}$} & \colhead{EM$_{1}$} &\colhead{kT$_{2}^{d}$} & \colhead{EM$_{2}$} & \colhead{Ne$^{e}$} &  \colhead{Norm$^{f}$} & \colhead{}  \\
\colhead{} & \colhead{} &\colhead{($10^{-2}cm^{-6}pc$)} & \colhead{(keV)} & \colhead{($10^{-2}cm^{-6}pc$)} &\colhead{(keV)} & \colhead{($10^{-2}cm^{-6}pc$)} & \colhead{}  &\colhead{} & \colhead{} & \colhead{} \\
} 
\rotate
\startdata 
{\bf $\rm 1T^{I}$}&&&&&&&& \\
\hline
{Off-field~2} &  $3.1\pm1.3$  &   $1.72\pm0.41$   & $0.198\pm0.007$ & $1.45\pm0.18$    &  \nodata & \nodata   &   \nodata     & $8.89\pm0.57$  & $1000.9/937 $ \\
{Off-field~3} & $1.2\pm0.8$  &$1.47\pm0.25$   & $"$     & $0.65\pm0.10$    &  \nodata    & \nodata     &   \nodata     & $13.17\pm0.47$  & $"$ \\
{Off-field~4} & $3.3\pm1.0$  &$2.00\pm0.30$   & $" $     & $0.92\pm0.15$    &  \nodata    & \nodata     &   \nodata   &   $12.57\pm0.55$  &     $"$ \\
{Off-field~5} & $0.4\pm0.7$  &$1.75\pm0.36$   & $"$     & $1.15\pm0.15$    &  \nodata    & \nodata     &   \nodata   &   $10.78\pm0.55$  &    $"$ \\
\hline
{\bf $\rm 2T^{II}$} &&&&&&&& \\
\hline
{Off-field~2}& $2.1\pm1.4$  & $1.51\pm0.47$   & $0.176\pm0.008$&  $1.80\pm0.29$   &  $0.703\pm0.117$    &  $0.10\pm0.04$ & \nodata & $8.27\pm0.62$  &  $939.3/929 $\\
{Off-field~3}& $1.0\pm0.8$  & $1.47\pm0.29$   & $"$   &  $0.78\pm0.16$   &  $0.806\pm0.089$    &  $0.09\pm0.02$ & \nodata & $12.49\pm0.54$  &  $"$\\
{Off-field~4}& $2.5\pm1.1$  & $1.82\pm0.34$   & $"$   &  $1.23\pm0.24$   &  $0.698\pm0.339$    &  $0.04\pm0.03$ & \nodata &  $12.55\pm0.61$  & $"$\\
{Off-field~5}& $0.0\pm0.0$  & $1.62\pm0.42$   & $" $   &  $1.39\pm0.21$   &  $0.715\pm0.088$    &  $0.09\pm0.03$ & \nodata &  $10.31\pm0.62$  & $"$\\
\hline
{\bf 1T$_{\scriptsize Ne}^{III}$} &&&&&&&& \\
\hline
{Off-field~2} & $3.0\pm1.3$  & $1.75\pm0.43$ & $0.195\pm0.007$ & $1.47\pm0.20$    &  \nodata & \nodata &   $1.9\pm0.5$    &   $8.71\pm0.58$  &     $956.3/933$ \\
{Off-field~3} & $1.3\pm0.8$  &  $1.52\pm0.26$ & $"$     & $0.65\pm0.11$    &  \nodata & \nodata &   $3.1\pm0.9$    &    $12.92\pm0.48$  &   $"$ \\
{Off-field~4} & $3.2\pm1.0$  & $1.98\pm0.30$ & $"$     & $0.95\pm0.16$    &  \nodata & \nodata &   $1.5\pm0.8$    &     $12.58\pm0.56$  &   $"$ \\
{Off-field~5} & $0.5\pm0.7$  & $1.80\pm0.37$ & $"$     & $1.16\pm0.16$    &  \nodata & \nodata &   $2.3\pm0.6$    &   $10.59\pm0.56$  & $"$ \\
\enddata
\tablecomments{\\
$^{a}$ Normalization of the Gaussian at the fixed center energy E=0.525 keV in units of L.U. ($\rm photons~s^{-1}~cm^{-2}~str^{-1}$).\\
$^{b}$ Emission measure for foreground (LB+SWCX) component with temperature fixed at kT=0.1 keV.\\
$^{c}$ Temperature of Galactic halo warm-hot phase thermal component.\\
$^{d}$ Temperature of Galactic halo hot phase thermal component.\\
$^{e}$ Ne abundance for Galactic halo warm-hot phase component.\\
$^{f}$ Normalization of the power-law model with fixed photon index $\rm \Gamma=1.52$ in the units of $\rm photons~keV^{-1}~s^{-1}~sr^{-1}~cm^{-2}$.\\
$^{I}$ Galactic halo one temperature model.\\
$^{II}$ Galactic halo two temperature model.\\
$^{III} $Galactic halo one temperature model with variable Ne abundance.
}
\end{deluxetable*}

%%%%%%%%%%%%%%%%%%%%%%%%%%%%%%%%%%%%%%%%%%%%%%%%%%%%%%%%%%%%%%%%%%%%%%%%%%%%%%%%%%%%%%%%%%%%%%%%%%%

%-----------------------------Figure Start------------------------------

\begin{figure*}
    \centering
    \includegraphics[scale=0.4]{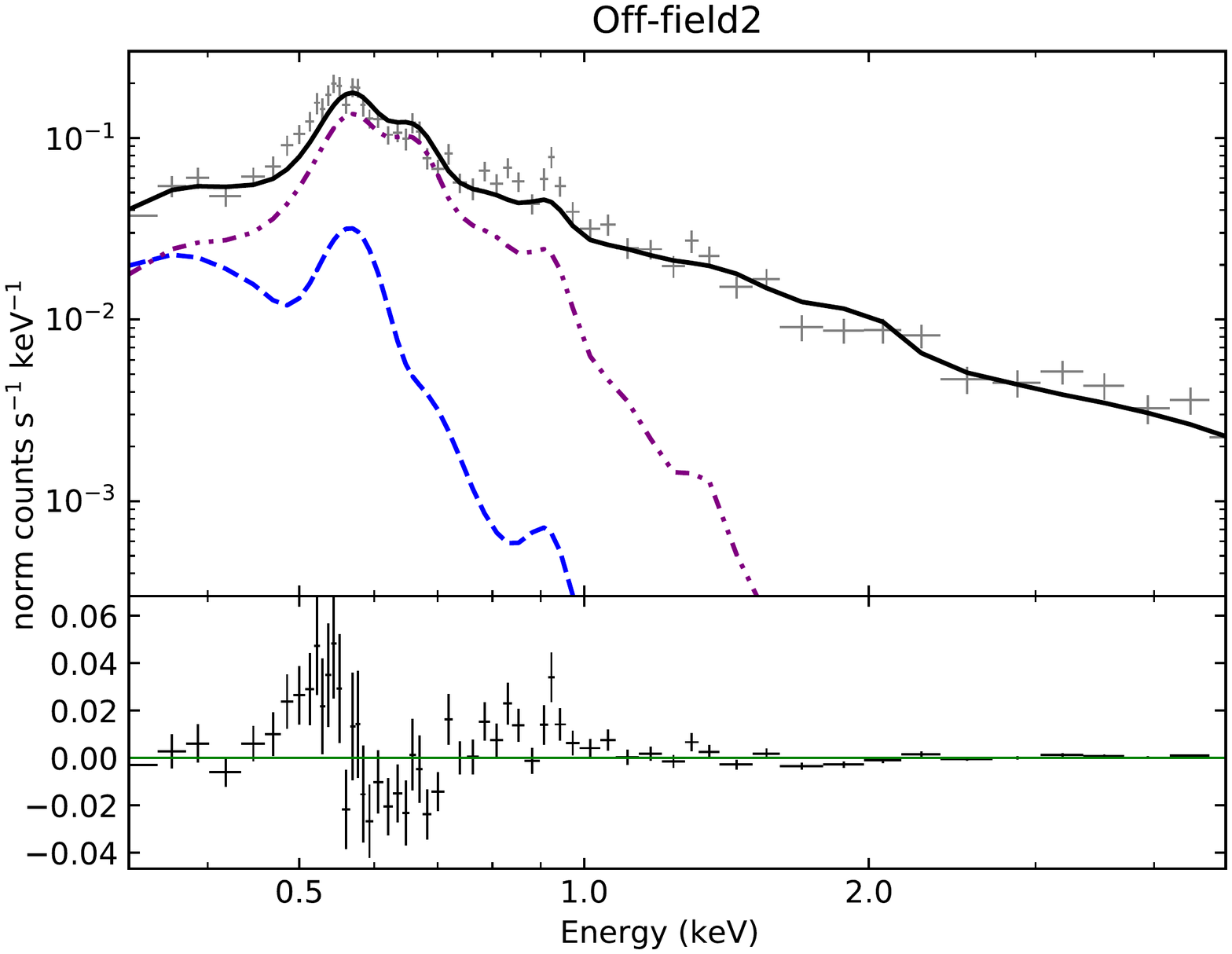}
    \includegraphics[scale=0.4]{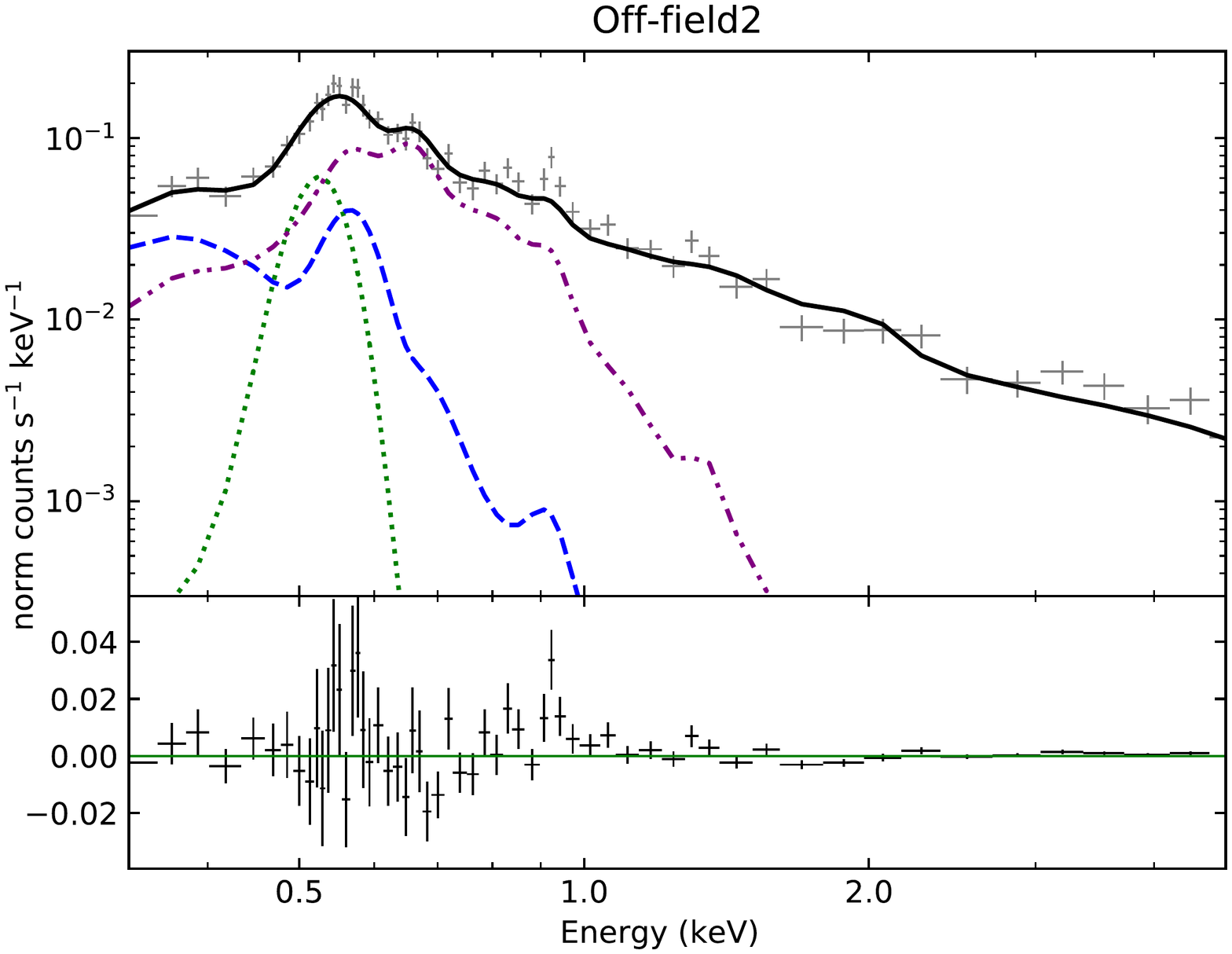}
    \vspace{-20 mm}
    \caption{{\it \bf Left}: \offii \suzaku XIS1 spectrum with the standard SDXB three component best fit model. Excess emission near $\rm 0.5~keV$ can be clearly seen in the residual plot. {\it \bf Right} SDXB three component model plus a Gaussian line accounting for the \oi emission. The blue curve shows the foreground spectrum and the purple curve shows the spectrum of the Milky Way CGM in the warm-hot phase. 
    }
    \label{fig:1}
\end{figure*}

%-----------------------------Figure end------------------------------

%-----------------------------Figure Start------------------------------

\begin{figure}
    \centering
    \vspace{-20 mm}
    \includegraphics[scale=0.4]{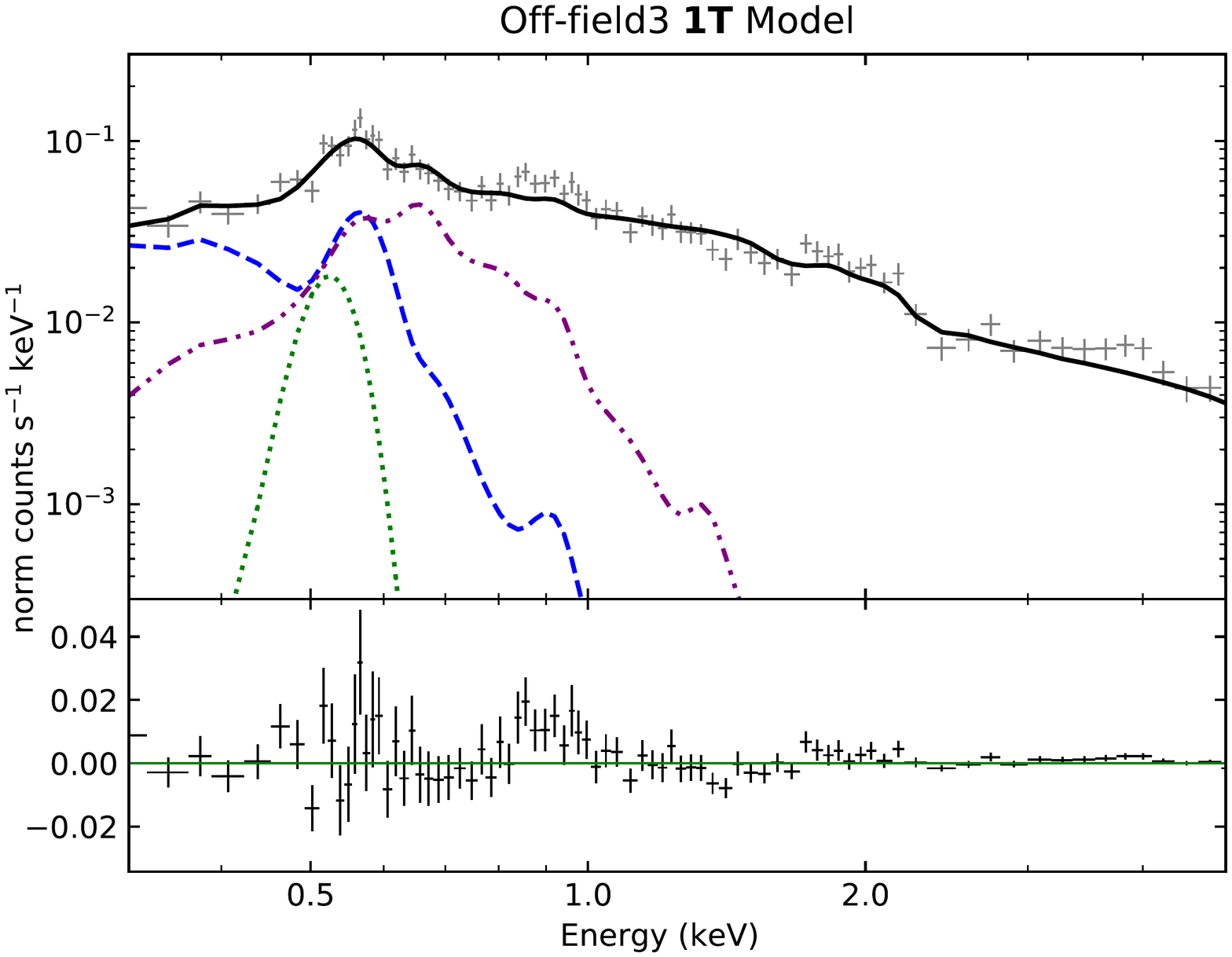} 
    
    \vspace{-40 mm}
    \includegraphics[scale=0.4]{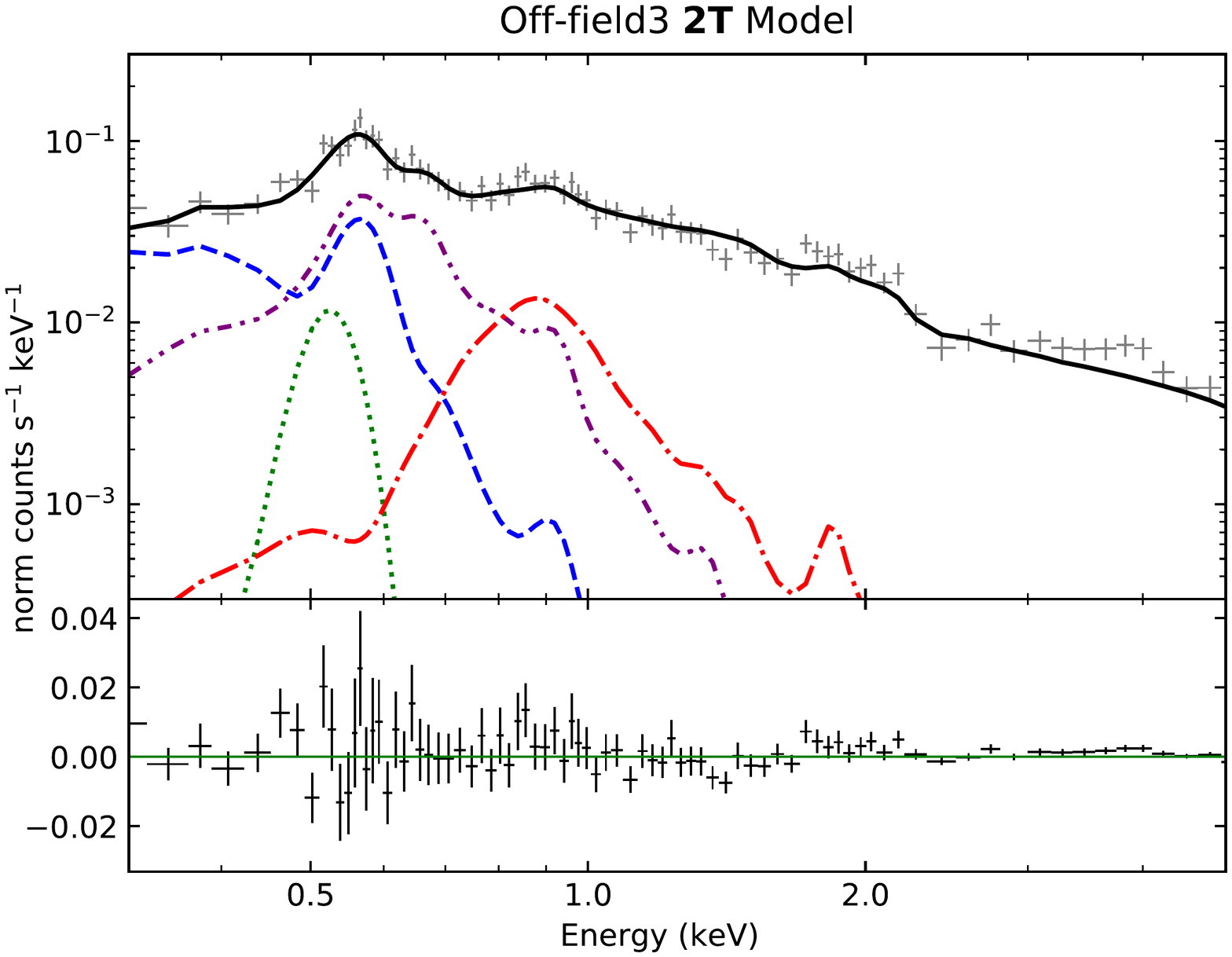} 
    
    \vspace{-40 mm}
    \includegraphics[scale=0.4]{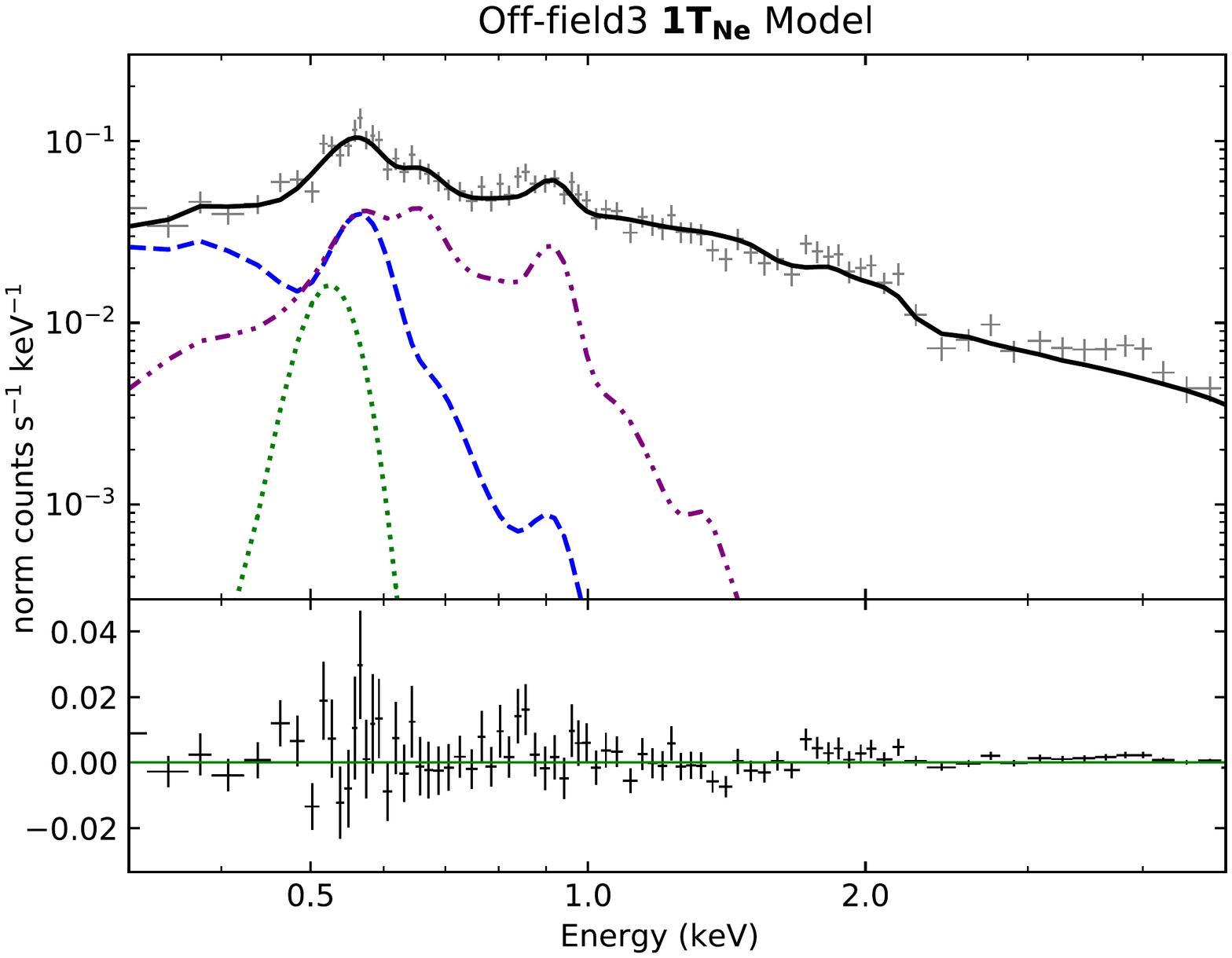} 
    \vspace*{-20mm}
\caption{Off-field 3 spectral fits. Top: data fitted with the 1T model; note the excess near $0.9$~keV. Middle: 2T model fit; the excess emission is modeled as an additional hot thermal component. Bottom: $\rm 2T_{Ne}$ model fit: the excess emission is modeled as a Ne emission line. 
}
\end{figure}

%-----------------------------Figure end------------------------------

%-----------------------------Figure Start------------------------------

\begin{figure*}
    \centering
    \includegraphics[scale=1.0]{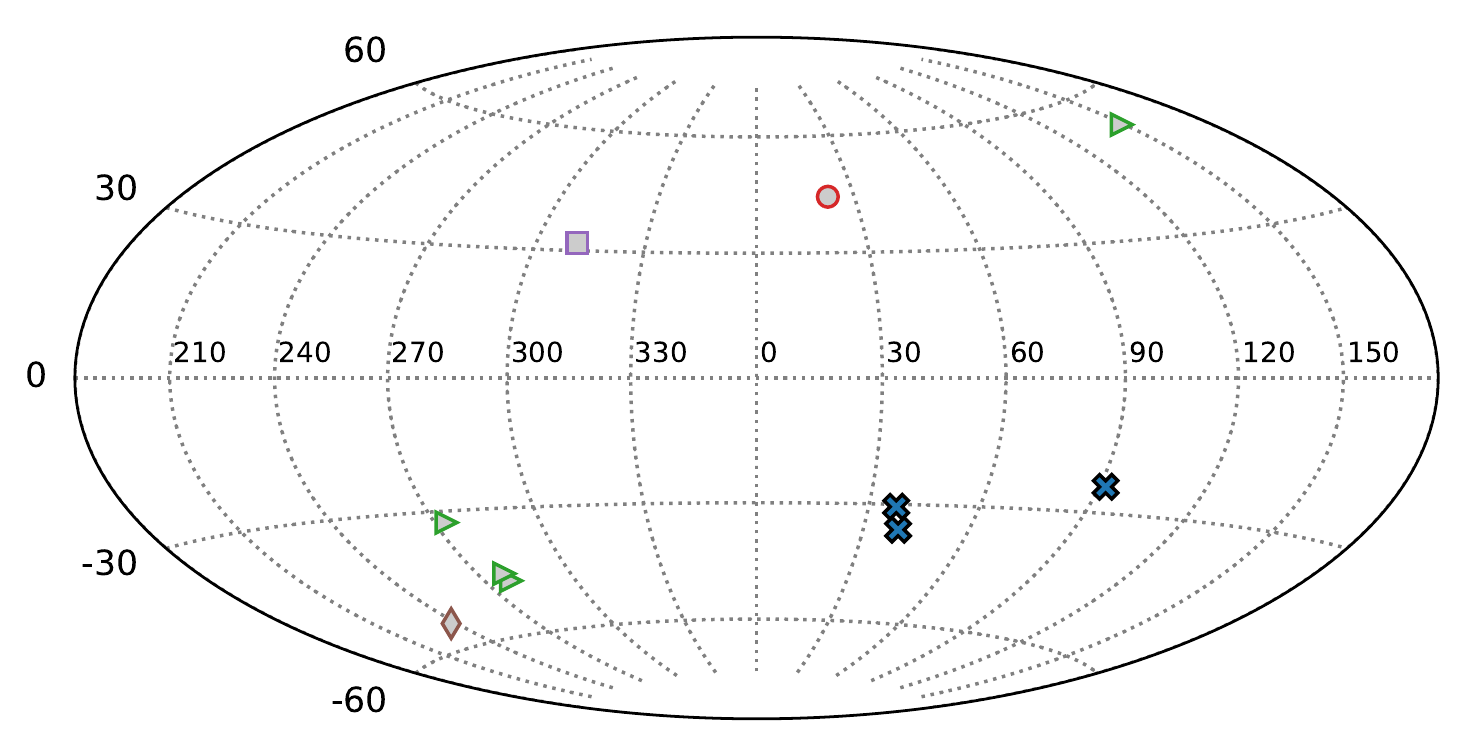} 
    \vspace{3 mm}
\caption{Sky map showing the locations of the sight lines with detections of the hot component of the CGM and/or Ne overabundant warm-hot gas. The sightlines shown with blue crosses are from this work, the red circle is from \citet{Das2019a, Das2019c}, the green triangles are from \cite{Yoshino2009}, the purple square is from \cite{Mitsuishi2012} and the brown diamond is from \cite{Henley2013}. The Galactic center is at the center of the plot. The figure shows that the presence of the hot component and/or Ne overabundance is not necessarily associated with the structures at the Galactic center, such as the Fermi Bubble. 
}
\end{figure*}

%-----------------------------Figure end------------------------------

\clearpage

\bibliographystyle{aasjournal}

\end{document}